\documentclass[11pt]{article}
\usepackage[margin=1in]{geometry}
\usepackage{natbib}
\usepackage{amsmath,amssymb,amsthm}
\usepackage[ruled,vlined,linesnumbered]{algorithm2e}
\usepackage[hyphens]{url}
\usepackage{hyperref}
\usepackage[utf8]{inputenc}
\usepackage[T1]{fontenc}
\usepackage{graphicx}
\usepackage{longtable}
\usepackage{booktabs}
\usepackage{array}
\usepackage{placeins}
\usepackage[most]{tcolorbox}
\newtcolorbox{appliedbox}[1][What this means for applied researchers]{%
  breakable, arc=1pt, boxrule=0.5pt,
  colframe=black!45, colback=black!4, colbacktitle=black!10, coltitle=black,
  fonttitle=\bfseries\small, fontupper=\small,
  left=8pt, right=8pt, top=6pt, bottom=6pt, boxsep=2pt, title={#1}}
\newenvironment{keywords}{\par\medskip\noindent\textbf{Key words:} }{\par\medskip}
\date{}

\newtheorem{theorem}{Theorem}
\newtheorem{proposition}[theorem]{Proposition}

\theoremstyle{definition}
\newtheorem{definition}[theorem]{Definition}
\theoremstyle{remark}
\newtheorem{remark}[theorem]{Remark}

\newcommand{\E}{\mathbb{E}}
\newcommand{\Var}{\mathrm{Var}}
\renewcommand{\Pr}{\mathbb{P}}
\newcommand{\R}{\mathbb{R}}
\newcommand{\W}{W}\newcommand{\A}{A}\newcommand{\Y}{Y}
\newcommand{\Ptil}{\widetilde{P}}\newcommand{\Phat}{\widehat{P}}
\newcommand{\Qbar}{\bar{Q}}\newcommand{\Qstar}{\Qbar^{*}}
\newcommand{\gzero}{g_{0}}
\newcommand{\psistar}{\psi^{*}}
\newcommand{\Mmod}{\mathcal{M}}\newcommand{\Mdelta}{\mathcal{M}_{\delta}}
\newcommand{\Dstar}{D^{*}}\newcommand{\expit}{\operatorname{expit}}
\newcommand{\TV}{\mathrm{TV}}\newcommand{\bstar}{\beta^{*}}
\newcommand{\indep}{\perp\!\!\!\perp}

\newcommand{\rhcN}{5{,}735}
\newcommand{\numReps}{1{,}000}\newcommand{\numCells}{16}\newcommand{\slV}{5}
\newcommand{\floorLo}{0.18}\newcommand{\floorHi}{0.00}\newcommand{\floorGT}{0.05}\newcommand{\psmin}{0.003}
\newcommand{\numBstar}{0.69}\newcommand{\numMargLOR}{0.62}\newcommand{\numMargRD}{0.13}
\newcommand{\covIpwST}{XX}\newcommand{\covTmleST}{XX}\newcommand{\covIpwGT}{XX}\newcommand{\covTmleGT}{XX}
\newcommand{\covGcompST}{XX}\newcommand{\covGcompGT}{XX}
\newcommand{\covTmleGlmST}{XX}
\newcommand{\covAipwSlST}{XX}\newcommand{\covAipwSlGT}{XX}
\newcommand{\covTmleCoarseST}{XX}
\newcommand{\biasTmleGbLoose}{XX}\newcommand{\biasTmleGbDef}{XX}\newcommand{\biasTmleGbTight}{XX}
\newcommand{\naTmleST}{XX}\newcommand{\naTmleGT}{XX}
\newcommand{\rmseGcompOK}{XX}\newcommand{\rmseAipwOK}{XX}\newcommand{\rmseTmleOK}{XX}
\newcommand{\rmseGcompWR}{XX}\newcommand{\rmseAipwWR}{XX}\newcommand{\rmseTmleWR}{XX}
\newcommand{\essLo}{XX}\newcommand{\essHi}{XX}
\newcommand{\covTmleCfInsST}{XX}\newcommand{\covAipwCfInsST}{XX}
\newcommand{\covTmleCfST}{XX}\newcommand{\covAipwCfST}{XX}
\newcommand{\covTmleCfGT}{XX}\newcommand{\covTmleCfKsix}{XX}
\newcommand{\biasTmleCfInsST}{XX}\newcommand{\biasTmleCfST}{XX}
\IfFileExists{sim_results.tex}{\renewcommand{\covGcompST}{0.95}
\renewcommand{\covGcompGT}{0.96}
\renewcommand{\covIpwST}{0.96}
\renewcommand{\covIpwGT}{0.95}
\renewcommand{\covTmleST}{0.29}
\renewcommand{\covTmleGT}{0.94}
\renewcommand{\covTmleGlmST}{0.93}
\renewcommand{\covAipwSlST}{0.58}
\renewcommand{\covAipwSlGT}{0.94}
\renewcommand{\covTmleCoarseST}{0.95}

\renewcommand{\naTmleST}{0\%}
\renewcommand{\naTmleGT}{0\%}
\renewcommand{\biasTmleGbTight}{0.008}
\renewcommand{\biasTmleGbDef}{0.028}
\renewcommand{\biasTmleGbLoose}{0.040}
\renewcommand{\rmseGcompOK}{0.013}
\renewcommand{\rmseAipwOK}{0.018}
\renewcommand{\rmseTmleOK}{0.016}
\renewcommand{\rmseGcompWR}{0.028}
\renewcommand{\rmseAipwWR}{0.019}
\renewcommand{\rmseTmleWR}{0.015}
\renewcommand{\essLo}{2,802}
\renewcommand{\essHi}{3,606}
}{}
\IfFileExists{crossfit_results.tex}{\renewcommand{\covTmleCfInsST}{0.31}
\renewcommand{\covAipwCfInsST}{0.59}
\renewcommand{\covTmleCfST}{0.50}
\renewcommand{\covAipwCfST}{0.80}
\renewcommand{\covTmleCfGT}{0.96}
\renewcommand{\covTmleCfKsix}{0.94}
\renewcommand{\biasTmleCfInsST}{0.027}
\renewcommand{\biasTmleCfST}{0.007}
}{}

\title{A Plasmode Benchmark Can Refute but Cannot Certify:\\
Projection and Regularity in Causal-Inference Simulation}
\author{M. Ehsan Karim\\[3pt]
  \normalsize School of Population and Public Health, University of British Columbia,\\
  \normalsize Vancouver, British Columbia, Canada;\\[1pt]
  \normalsize and Centre for Advancing Health Outcomes, St.\ Paul's Hospital,\\
  \normalsize Vancouver, British Columbia, Canada\\[2pt]
  \normalsize \texttt{ehsan.karim@ubc.ca}\\[1pt]
  \normalsize ORCID: \href{https://orcid.org/0000-0002-0346-2871}{0000-0002-0346-2871}}

\begin{document}
\maketitle

\begin{abstract}
\noindent Plasmode simulation---resampling covariates, and often treatment, from a
real cohort and regenerating the outcome from a model with an injected effect---is a
standard tool for comparing causal-inference estimators. We argue that its validity
turns on two separable design choices, and that separating them clarifies a recent
debate over resampling versus regenerating the treatment. Our central message limits
what such a benchmark can show: a plasmode can \emph{refute} a universal claim about an
estimator but cannot \emph{certify} one. Refuting needs only that the simulated data
genuinely belong to the class over which a guarantee is quoted; certifying also needs
the benchmark's difficulty to carry over to the target population, which the analyst
cannot verify. Two choices set that difficulty. \emph{Regularity} is whether the
simulated law preserves the overlap that inverse-weighting and targeted-learning theory
require. Resampling the real treatment with fine-grained covariates destroys it, and the
resulting coverage failure falls on flexible, machine-learning-based estimators, while
parametric-nuisance estimators---including inverse-probability weighting---stay near
nominal; it disappears once the treatment is regenerated from a bounded propensity or
the covariates are coarsened, and cross-fitting removes only part of it.
\emph{Projection} is that the injected outcome model hands a matching estimator the
right answer by construction; this survives the regularity fix and can reverse estimator
rankings even when overlap is intact. We give a regular-plasmode algorithm, a confirming
real-data simulation, and practical recommendations. A plasmode is a filter against
estimators, not a warrant for them.
\end{abstract}


\section{Introduction}\label{sec:intro}

\textbf{Background.} Plasmode simulation evaluates a causal estimator by resampling
covariates---and often the treatment---from a real cohort, then regenerating the
outcome from a model into which a known effect has been injected
\citep{cattell1967plasmode,gadbury2008,vaughan2009,franklin2014plasmode}. The appeal
is that the covariate law is real, so the benchmark inherits the correlation
structure, dimensionality, and near-violations that computer-generated designs omit,
while the effect is known because the analyst placed it there. The construction is
standard in pharmacoepidemiology and is packaged for reuse
\citep{franklin2017plasmode,schreck2024}. Two treatment mechanisms are in use: under
\emph{sample-treatment} the real treatment is carried through the resampling and only
the outcome is regenerated; under \emph{generate-treatment} the treatment is also
regenerated from a fitted propensity \citep{franklin2014plasmode,schreck2024}.
\citet{shaw2026} recently argued that the sample-treatment mechanism induces a
positivity violation that biases the measured performance of propensity-based
estimators, and recommended generate-treatment on those grounds.

\textbf{The estimand.} Throughout, the target is the marginal average treatment
effect on the risk-difference scale for a binary outcome,
\begin{equation}
\psi(P) = \E_{P}\!\big[\Qbar_P(1,\W)-\Qbar_P(0,\W)\big],
\qquad \Qbar_P(a,w)=\E_P[\Y\mid \A=a,\W=w],
\label{eq:estimand}
\end{equation}
identified from $(\W,\A,\Y)$ under consistency, exchangeability $\Y(a)\indep\A\mid\W$,
and positivity $0<g(\W)<1$, with $g(w)=\Pr(\A=1\mid\W=w)$ the propensity;
\eqref{eq:estimand} is the standardization (g-)formula \citep{robins1986}, the target
of inverse probability weighting (IPW), augmented IPW (AIPW), and targeted
minimum-loss estimation (TMLE).

\textbf{Contributions.} We recast the validity of a data-anchored simulation benchmark as
two separable design axes---\emph{projection} and \emph{regularity}---and ask what such a
benchmark can and cannot establish about an estimator. (i) We show the sample-treatment
positivity failure is a \emph{regularity} statement---the marginal effect is not pathwise-differentiable at
the simulated law relative to the nonparametric model $\Mmod$, so the law lies outside
the strict-overlap submodel $\Mdelta$ in which IPW/AIPW/TMLE guarantees hold---indexed by covariate granularity, so
the high-dimensional near-degeneracy of \citet{damour2021} is realized exactly and at
finite $n$ by resampling, and absent under coarse covariates
(Section~\ref{sec:regularity}); this sharpens \citet{shaw2026} and reconciles it with
its coarse-covariate counterexample. (ii) We show the injected outcome model is a
\emph{projection}: it fixes the truth exactly but advantages an estimator whose
working model matches it, so the estimator ranking measured on a plasmode need not
transfer, and this failure is untouched by the generate-treatment fix
(Section~\ref{sec:projection}). (iii) We prove the resulting asymmetry: an anchor
refutes a universal reliability claim with no transfer assumption, but certifies none
(Section~\ref{sec:conclude}). (iv) We give an \emph{algorithm} for a regular plasmode
that repairs the regularity failure by construction, recomputes the truth by
standardization, and declares its difficulty coordinates (Section~\ref{sec:algo}),
and (v) a simulation study spanning conventional and super-learner-based estimators
on real data that exhibits both failures, the repair, and a difficulty dial
(Section~\ref{sec:sim}). Proofs are in Web Appendix~A, which also tabulates the status and governing
model of each result (Web Table~S1); the software is provided. Table~\ref{tab:roadmap} maps these results to the questions they settle and
to where each is stated and proved.

\textbf{Scope.} Results are stated for the binary-outcome logistic plasmode that
matches deployed software \citep{franklin2017plasmode}. All statements of
irregularity are relative to the nonparametric model $\Mmod$; within the injected
parametric outcome model the same functional is smooth with a finite bound, and we
mark this reading wherever it matters. We do not treat time-varying treatment or
censoring, and we do not give a data-computable test of whether a ranking transfers
to a named population---that gap is the one open problem.

\begin{table}[t]\centering\small
\caption{Map of the results. Each row names a result, the question it settles, and
where its formal statement and proof appear. ``Model'' indicates whether the statement
is relative to the nonparametric model $\Mmod$ or holds within the injected parametric
outcome model.}
\label{tab:roadmap}
\begin{tabular}{@{}p{0.30\textwidth} p{0.44\textwidth} p{0.18\textwidth}@{}}
\toprule
Result & Question it settles & Location \\
\midrule
Projection (Props.~\ref{prop:noncollapse},~\ref{prop:rank}) & The injected model is a
projection: a matched working model is correct by construction, so the plug-in-vs-
doubly-robust ranking need not transfer. & \S\ref{sec:projection}; Web App.~A \\
\addlinespace
Regularity (Thm.~\ref{thm:reg}) & Sample-treatment positivity failure is
granularity-indexed: exact under near-unique covariates, absent under coarse ones.
Relative to $\Mmod$. & \S\ref{sec:regularity}; Web App.~A \\
\addlinespace
No-free-lunch (Rem.~\ref{rem:vonmises}) & Distributional fidelity bounds the estimand
gap, not the estimator ranking. & \S\ref{sec:together} \\
\addlinespace
Refute/certify (Thm.~\ref{thm:falsify}) & Refutation needs no transfer assumption;
certification provably requires an unverifiable transfer premise (T2 the canonical instance). &
\S\ref{sec:conclude}; Web App.~A \\
\addlinespace
Regular anchor (Prop.~\ref{prop:regular}) & A declared overlap floor plus a recomputed
truth restores an in-model, regular target. & \S\ref{sec:algo}; Web App.~A \\
\bottomrule
\end{tabular}
\end{table}

\section{The plasmode data-generating process and the model}\label{sec:setup}

Observations are $O=(\W,\A,\Y)$ with covariates $\W\in\R^{p}$, binary $\A\in\{0,1\}$,
binary $\Y\in\{0,1\}$, from a source law $P_{0}$. For a law $P$ write
$g_P(w)=\Pr_P(\A=1\mid\W=w)$ and $\Qbar_P(a,w)=\E_P[\Y\mid\A=a,\W=w]$, and abbreviate
$\gzero=g_{P_0}$. Let $\Mmod$ be the nonparametric model and $\Mdelta\subset\Mmod$ the
submodel of laws with strict overlap $\delta\le g_P(w)\le 1-\delta$, $\delta>0$. On
$\Mdelta$ the functional $\psi$ of \eqref{eq:estimand} is pathwise differentiable with
efficient influence function
\begin{equation}
\Dstar_{P}(O)=\frac{\A}{g_P(\W)}\{\Y-\Qbar_P(1,\W)\}
-\frac{1-\A}{1-g_P(\W)}\{\Y-\Qbar_P(0,\W)\}+\Qbar_P(1,\W)-\Qbar_P(0,\W)-\psi(P),
\label{eq:eif}
\end{equation}
and efficiency bound $\Var_P\Dstar_P$, both finite under strict overlap---and more
generally whenever $\Dstar_P\in L_2(P)$, which strict overlap ensures.

A plasmode operator fits a parametric outcome model, in the deployed binary case a
logistic regression, \emph{injects} a treatment coefficient $\bstar$, optionally
rescales the confounder coefficients, and re-solves the intercept to a target event
rate, giving the modified regression
\begin{equation}
\Qstar(a,w)=\expit(\alpha_0^{*}+\bstar a+w^{\top}\gamma^{*}).
\label{eq:qstar}
\end{equation}
It then draws a covariate--treatment configuration by resampling and regenerates
$\Y^{\#}\mid(\A,\W)\sim\mathrm{Bernoulli}(\Qstar(\A,\W))$.

\begin{definition}[Sample- and generate-treatment plasmode]\label{def:st-gt}
\emph{Sample-treatment} draws $(\W,\A)$ from the empirical joint $\Phat_n$ and sets
$\Y^{\#}$ from \eqref{eq:qstar}. \emph{Generate-treatment} draws $\W$ from the
empirical marginal, draws $\A^{\#}\sim g(\cdot\mid\W)$ from a propensity, and sets
$\Y^{\#}$ from \eqref{eq:qstar}. Write $\Ptil_n$ for the resulting law.
\end{definition}

\section{The projection axis}\label{sec:projection}

Because the analyst sets \eqref{eq:qstar}, the benchmark's truth is a functional of a
chosen object: the marginal risk difference $\psistar=\E_{\widetilde\W}[\Qstar(1,\widetilde\W)-\Qstar(0,\widetilde\W)]$
over the realized simulated covariate law. It is computable to arbitrary Monte-Carlo
precision regardless of overlap; where strict overlap holds it equals the
nonparametrically identified $\psi(\Ptil_n)$ (Web Appendix~A). The number the
analyst types, however, is not the marginal target.

\begin{proposition}[Scale and non-collapsibility]\label{prop:noncollapse}
In the logistic plasmode \eqref{eq:qstar} with no treatment--covariate interaction,
$\bstar$ is the constant conditional log-odds-ratio. The marginal risk difference
$\psistar$ and the marginal log-odds-ratio both differ from $\bstar$---the first by
scale, the second by non-collapsibility of the odds ratio \citep{grp1999}, which
attenuates it strictly toward the null whenever the covariate contributes variance. A
marginal estimator must be scored against $\psistar$ recomputed by standardization,
not against $\bstar$.
\end{proposition}

On the illustration of Section~\ref{sec:sim}, an injected $\bstar=\log 2=\numBstar$
becomes a marginal log-odds-ratio of $\numMargLOR$ and a marginal risk difference of
$\numMargRD$ over the real covariate distribution; scoring an IPW, AIPW, or TMLE
estimate against $\numBstar$ would report a large bias that is an accounting error.

\begin{proposition}[Projection limits ranking transfer, not measurement validity]\label{prop:rank}
Wherever strict overlap holds, running AIPW or TMLE on plasmode replicates is an
internally valid measurement of their behavior at $\Ptil_n$. But an estimator that
takes the generating class \eqref{eq:qstar} as its working outcome model is correctly
specified, and its plug-in attains that model's information bound, no larger than the
nonparametric bound $\Var\Dstar$; on such a world this plug-in is \emph{asymptotically}
at least as efficient as a doubly-robust estimator, and---as the simulation of
Section~\ref{sec:sim} shows---its finite-sample error can match or beat one. The
measured ranking therefore need not transfer to a law where the working model is wrong. It does \emph{not} follow that a plasmode can assess only parametric
estimators, nor that conclusions about AIPW or TMLE at $\Ptil_n$ are unfounded.
\end{proposition}

The consequence, made concrete in Section~\ref{sec:sim}, is that even in a regular
generate-treatment world---where overlap is not at issue---a correctly specified
g-computation estimator has the smallest error when its outcome model matches the
generating one, and loses to AIPW and TMLE once the outcome surface carries structure
its model omits.

\begin{appliedbox}
When you build a plasmode you write down a formula for how the outcome depends on
treatment and covariates and inject a known effect into it. Any estimator whose own
model has that same shape is then correct \emph{by construction} and will tend to post
the smallest error on your benchmark---not because it is the better method, but because
you handed it the answer key. So the estimator \emph{ranking} a plasmode reports need
not be the ranking you would see on your real data, where no method has the answer key.
This is a caution about reading too much into who ``wins,'' not a claim that a plasmode
can only judge simple estimators: flexible methods such as AIPW and TMLE are measured
just as validly, wherever overlap holds. And the caveat is untouched by how you generate
the treatment---it is baked into the outcome model, so the generate-treatment fix
(Section~\ref{sec:algo}) does not remove it.
\end{appliedbox}

\section{The regularity axis}\label{sec:regularity}

\begin{theorem}[Granularity-indexed regularity breakdown]\label{thm:reg}
Take a sample-treatment plasmode with covariate patterns near-unique in the source,
so that with probability approaching one a resampled pattern carries a single
treatment value; there the simulated propensity lies in $\{0,1\}$, strict overlap
fails on a set of probability tending to one, and one arm's outcome regression is
not identified (the treated arm where the propensity is $0$, the control arm where it is $1$). Relative to $\Mmod$ the marginal effect \eqref{eq:estimand} is then
not pathwise differentiable at $\Ptil_n$: no square-integrable gradient represents its
derivative, no efficient influence function exists, the efficiency bound is $+\infty$ (made precise as
divergence along identified interior sequences approaching $\Ptil_n$; Web Appendix~A),
and $\Ptil_n\notin\Mdelta$ for every $\delta>0$. With coarse covariates whose strata
each contain both treatments, overlap holds and $\psi$ is regular.
\end{theorem}

\textbf{Reconciliation.} Theorem~\ref{thm:reg} settles when the sample-treatment
positivity violation holds. \citet{shaw2026} are right that sample-treatment produces
a violation in the regime plasmode is built for; the theorem adds that a
coarse-covariate case is regular, so the violation is granularity-indexed rather than
universal. The population geometry is not ours: \citet{damour2021} show that strict
overlap becomes increasingly restrictive as informative high-dimensional covariates
accumulate; sample-treatment realizes that tension exactly and at finite $n$ by freezing
the propensity at the source treated-fractions. On the right-heart-catheterization
cohort (Section~\ref{sec:sim}) the overlap floor collapses from \floorLo{} at coarse
strata to \floorHi{} once the real covariates are resolved, against a floor of
\floorGT{} that a generate-treatment design can declare; the cohort's own fitted
propensity dips to \psmin{}, so a declared floor is a design choice, not a loss of
realism. The near-unique regime is not a corner case: it is the regime plasmode
advertises, so realism and regularity are in tension.

\begin{appliedbox}
If you keep the real treatment when you resample (sample-treatment) and your covariates
are detailed enough that almost every patient's covariate pattern is unique, then each
pattern carries only the one treatment that patient actually received. The benchmark's
propensity is then effectively $0$ or $1$, and the overlap condition that inverse-weighting
and targeted learning rely on is gone. On the right-heart cohort this is exactly what
happens at full covariate resolution, and it drives the flexible, super-learner versions
of TMLE and AIPW well below their nominal coverage---while the same estimators behave once
you either coarsen the covariates or, better, generate the treatment from a propensity you
bound away from $0$ and $1$ (Section~\ref{sec:algo}). The simulation
(Section~\ref{sec:sim}) shows both the lost overlap and the way these flexible methods are
fitted contribute to the gap. Two cautions. This is a property of \emph{how detailed} your
covariates are, not a blanket verdict on sample-treatment---coarse designs stay regular.
And the simpler, parametric estimators that keep their coverage here do so in part because
their rigid model cannot resolve the near-determinism---not because the benchmark is sound;
do not read their good coverage as reassurance.
\end{appliedbox}

\section{The two axes together: a design no-free-lunch}\label{sec:together}

\textbf{Four requirements, and what each family gives up.} A benchmark with a known
truth must specify the component that carries the effect, and that choice trades off
four things: (i) nonparametric fidelity to the real law; (ii) a regular target with
strict overlap and an existing influence function; (iii) no fabricated treatment law;
and (iv) an external truth pinned to the source rather than to a fitted generator.

\textbf{No design has all four.} Each established family sacrifices at least one of
(i)--(iv). Sample-treatment plasmode keeps (i) and (iii) but fails (ii) in the
near-unique regime (Theorem~\ref{thm:reg}). Generate-treatment plasmode restores (ii)
only by fabricating a parametric propensity, departing from (i) and (iii) on the
treatment law. Generative designs \citep{parikh2022,athey2024,neal2020} can attain
(i)--(iii) internally but pin the truth---and, more consequentially, the difficulty an
estimator faces---to a learned generator rather than to $P_0$, failing (iv). This is a
per-family tradeoff, not an impossibility; it places the existing designs as cells of
the projection--regularity grid.

\textbf{Fidelity does not rescue the ranking.} One might hope a high-fidelity simulator
inherits the source's estimator ranking. It does not, and the reason is the same
overlap floor.

\begin{remark}[Fidelity bounds the estimand gap, not the ranking]\label{rem:vonmises}
A first-order expansion gives
$|\psi(Q)-\psi(P)|\le\|\Dstar_P\|_\infty\,2\,\TV(P,Q)+|R_2(P,Q)|$, where
$\TV(P,Q)=\sup_{B}|P(B)-Q(B)|$ is the total-variation distance and
$\|\Dstar_P\|_\infty$ is of order $1/\delta$ at overlap floor $\delta$. The linear term
is controlled by total-variation fidelity, and on $\Mdelta$ the remainder is the
doubly-robust bilinear term
$R_2(P,Q)=O\!\big(\delta^{-1}\!\int|g_P-g_Q|\,|\Qbar_P-\Qbar_Q|\,dQ\big)=O(\delta^{-1}\TV(P,Q))$
for $\Y\in[0,1]$, so the \emph{estimand} gap is $O(\delta^{-1}\TV(P,Q))$---controlled by
fidelity, but degrading as $1/\delta$ and vacuous at the positivity boundary. Either way it says nothing about the
\emph{difficulty} an estimator faces or the resulting ranking, which are what a
benchmark reports; that the ranking need not follow is interpretive, not a corollary of
the inequality. The $1/\delta$ blow-up here is the same overlap floor that
Theorem~\ref{thm:reg} drives to zero.
\end{remark}

\begin{appliedbox}
No plasmode design in current use is realistic, keeps overlap intact, invents no treatment
mechanism, and pins its truth to the real cohort all at once; each established design gives
up at least one---a tradeoff across the designs in use, not a proven impossibility for
every conceivable one---so choosing a design means choosing which compromise you can live
with. One tempting escape does not work: making the simulator more \emph{realistic}. A
closer match to the real data limits how far the benchmark's true effect can drift from the
real one---though even that control weakens as overlap thins toward the boundary---but it says nothing about which estimator the benchmark makes look hardest or
best. Fidelity controls the target, not the difficulty an estimator faces or the ranking it
produces---so ``our generator is very realistic'' is not, by itself, a reason to trust the
estimator ordering it reports.
\end{appliedbox}

\section{A regular-plasmode algorithm}\label{sec:algo}

The regularity failure is removable by construction, and the same construction lets
the analyst \emph{declare} the difficulty at which estimators are compared rather than
read it off a fit. Algorithm~\ref{alg:regular} states the procedure.

\begin{algorithm}[t]
\SetAlgoLined\DontPrintSemicolon
\KwIn{source cohort $\{(\W_i,\A_i,\Y_i)\}_{i=1}^n$; injected conditional effect
$\bstar$; overlap floor $\delta\in(0,\tfrac12)$; declared difficulty coordinates
(confounder-strength multiplier $\eta$, effect-modification and nonlinearity terms);
target event rate $r$; replicate count $J$.}
\KwOut{$J$ simulated data sets and the exact marginal risk-difference truth $\psistar$.}
Fit $\widehat{\Qbar}(a,w)=\expit(\widehat\alpha_0+\widehat\beta a+w^{\top}\widehat\gamma)$
and a propensity $\widehat g(w)$ on the source cohort\;
Form the modified surface $\Qstar(a,w)=\expit(\alpha_0^{*}+\bstar a+\eta\,w^{\top}\widehat\gamma+\text{declared terms})$;
solve $\alpha_0^{*}$ so the marginal event rate is $r$ under the source treatment assignment\;
\emph{(truth)} $\psistar \leftarrow$ exact empirical (g-formula) standardization over the source covariates
$\frac1n\sum_i\{\Qstar(1,\W_i)-\Qstar(0,\W_i)\}$\;
Bound the propensity: $\widetilde g(w)\leftarrow\min\{\max\{\widehat g(w),\delta\},1-\delta\}$\;
\For{$j\leftarrow 1$ \KwTo $J$}{
  Resample covariate rows $\W^{(j)}$ from the source with replacement\;
  Draw $\A^{(j)}\sim\mathrm{Bernoulli}(\widetilde g(\W^{(j)}))$
  \tcp*{generate-treatment: overlap $\ge\delta$ by construction}
  Draw $\Y^{(j)}\sim\mathrm{Bernoulli}(\Qstar(\A^{(j)},\W^{(j)}))$\;
}
Report, alongside each benchmark, the declared difficulty coordinates and the realized
overlap summaries (e.g.\ effective sample size of the weights)\;
\Return{$\{(\W^{(j)},\A^{(j)},\Y^{(j)})\}_{j=1}^J,\ \psistar$}\;
\caption{Regular plasmode with a declared overlap floor and difficulty coordinates.}
\label{alg:regular}
\end{algorithm}

The construction has one genuine cost, stated in Proposition~\ref{prop:regular}: it
occupies the generate-treatment corner---one fabricated but declared treatment
nuisance---and it certifies internal validity of the measurement only, never that the
ranking transfers. What it buys is that each hidden artifact of an ordinary plasmode
becomes a declared, inspectable coordinate: the analyst who imposes a floor $\delta$,
recomputes the truth by standardization (removing the scoring error of
Proposition~\ref{prop:noncollapse}), and reports the difficulty coordinates is
measuring behavior at a law the estimators' theory covers, and is stating rather than
concealing the difficulty at which the comparison was run.

\begin{proposition}[Declared overlap floor yields a regular, in-model anchor]\label{prop:regular}
For a plasmode built with treatment regenerated from a propensity bounded in
$[\delta,1-\delta]$, $\delta>0$, the marginal effect is pathwise differentiable at the
induced law, its efficient influence function exists, and the efficiency bound is
finite, so the induced law lies in $\Mdelta$---conditional on the source cohort and the
estimated propensity $\widehat g$ from which $\widetilde g$ is bounded. The floor is a
declared difficulty coordinate for the benchmark, not a claim of $\delta$-overlap in the
source population; the construction certifies internal validity only and does not
certify ranking transfer.
\end{proposition}

\section{Simulation study}\label{sec:sim}

\textbf{Design.} We follow the ADEMP structure \citep{morris2019}; Table~\ref{tab:design} summarizes the data-generating grid and the ten-arm estimator roster.

\begin{table}[!t]\centering\small
\caption{Simulation design. \emph{Panel A}: the reference-anchored data-generating grid---each axis is varied singly from the reference level (\textbf{bold}) rather than fully crossed, and outcome-model correctness is crossed with every cell. \emph{Panel B}: the ten estimator arms, each scored against the standardized truth. Held fixed across cells: $n=\rhcN$, injected treatment main-effect coefficient $\bstar=\log 2$ (the constant conditional log-odds-ratio only in outcome-correct cells), event rate calibrated to $0.5$ under the source treatment assignment, $J=\numReps$ replicates.}
\label{tab:design}
\begin{tabular}{@{}p{0.22\textwidth} p{0.40\textwidth} p{0.30\textwidth}@{}}
\toprule
\multicolumn{3}{@{}l}{\emph{Panel A. Data-generating grid --- reference-anchored, one axis varied at a time from the reference (bold)}}\\
\addlinespace
Factor & Levels & Role in the design \\
\cmidrule(r){1-1}\cmidrule(lr){2-2}\cmidrule(l){3-3}
Treatment mechanism & \textbf{generate-treatment}, sample-treatment & Regularity axis \\
Overlap floor $\delta$ & $0.01,\ 0.02,\ \mathbf{0.05},\ 0.10,\ 0.20$ & Difficulty dial \\
Confounder strength $\eta$ & $0.5,\ \mathbf{1},\ 1.5$ & Confounding axis \\
Outcome working model & \textbf{correct} (additive); wrong (omitted effect modifier $+$ nonlinearity) & Projection axis; crossed with every cell \\
Granularity $K$ (sweep) & \textbf{full} (reference), then $45,\ 23,\ 11,\ 6$ strata & Resolution probe (both mechanisms, outcome-correct) \\
\addlinespace
\multicolumn{3}{@{}p{0.92\textwidth}}{\footnotesize Reference cell: generate-treatment, $\delta=0.05$, $\eta=1$, full resolution, outcome model correct. Cell count: $(5\,\delta+2\,\eta+1\,\text{treatment})\times\{\text{correct},\,\text{wrong}\}=\numCells$ main cells; the granularity sweep adds the four coarsened resolutions $\times\{\text{ST},\text{GT}\}=8$, for $24$ stored cells. An estimator-free \emph{oracle} pass reads the true overlap floor and single-treatment mass off the construction at each $K$.}\\
\midrule
\multicolumn{3}{@{}l}{\emph{Panel B. Estimator arms --- ten arms scored against the standardized truth}}\\
\addlinespace
Arm (figure label) & Nuisance models & Type \\
\cmidrule(r){1-1}\cmidrule(lr){2-2}\cmidrule(l){3-3}
\multicolumn{3}{@{}l}{\emph{Primary comparison (7)}}\\
Unadjusted & --- & Naive reference \\
g-computation & Main-effects logistic (outcome) & Outcome plug-in \\
Subclassification & Main-effects logistic propensity & Propensity-score subclassification for the ATE (ten subclasses) \\
IPW & Main-effects logistic propensity, stabilized & Inverse-probability weighting \\
AIPW & Main-effects logistic & Doubly robust, parametric nuisances \\
AIPW-SL & Super learner & Doubly robust, flexible nuisances \\
TMLE & Super learner & Doubly robust, flexible nuisances \\
\addlinespace
\multicolumn{3}{@{}l}{\emph{Diagnostic TMLE variants (3)}}\\
TMLE-glm & Main-effects logistic & Bridge: same targeting step, no super learner \\
TMLE-gb.01, TMLE-gb.10 & Super learner; $g_{\min}\in\{0.01,0.025,0.10\}$ & Propensity-truncation sweep about the default $g_{\min}=0.025$ \\
\bottomrule
\end{tabular}

\smallskip
\footnotesize\raggedright
Super learner (both flexible arms, following \citealp{phillips2023}): $V=\slV$ folds stratified on the outcome for the outcome model and on treatment for the propensity model, negative-log-likelihood metalearner, library $=$ \{marginal mean; main-effects and elastic-net logistic regression; degree-$2$ MARS; random forest; BART\}. Full construction and the complete cell-by-estimator results are in Web Appendix~B (Web Table~S2).
\end{table}

\textbf{Aims.} To exhibit (a) the regularity failure of sample-treatment and its repair by a declared overlap floor; (b) that this failure is not merely a weak-learner artifact, locating the induced \emph{law}'s contribution versus the inferential procedure's (via cross-fitting); (c) the projection failure that survives the repair; and (d) the difficulty dial.

\textbf{Data-generating mechanisms.} Every cell shares the outcome-surface construction of Algorithm~\ref{alg:regular} on the right-heart-catheterization cohort \citep{connors1996} ($n=\rhcN$; treatment the receipt of a Swan--Ganz catheter, outcome in-hospital death, $50$ baseline covariates), with injected treatment main-effect coefficient $\bstar=\log 2$ and an intercept calibrated so the marginal event rate is $0.5$ under the source treatment assignment; the two mechanisms differ only in the treatment step---generate-treatment draws treatment from the bounded propensity of Algorithm~\ref{alg:regular}, whereas sample-treatment carries the source treatment forward with each resampled row. The grid is \emph{anchored at a reference difficulty} (generate-treatment; declared floor $\delta=0.05$; confounder-strength multiplier $\eta=1$), and varies each difficulty coordinate singly about that reference rather than fully crossing the axes, while outcome-model correctness (additive vs.\ an effect modifier plus a nonlinearity the analyst's working model omits) is crossed with every cell (Table~\ref{tab:design}A); this gives $(5{+}2{+}1)\times2=\numCells$ cells, and a granularity sweep over coarser covariate resolutions adds a further $8$, for $24$ stored cells (Web Appendix~B).

\textbf{Estimand.} The marginal risk difference \eqref{eq:estimand}, recomputed by standardization from each mechanism.

\textbf{Methods.} Ten estimator arms are scored against the standardized truth (Table~\ref{tab:design}B), in three tiers: five \emph{conventional} arms with main-effects logistic nuisances, two \emph{flexible} doubly-robust arms---AIPW-SL and TMLE---whose outcome and propensity nuisances come from a super learner, and three diagnostic TMLE variants. In the figures and Web Table~S2, the parametric AIPW is distinct from its super-learner counterpart AIPW-SL, and the conventional propensity-score subclassification arm for the ATE (ten subclasses) appears as \emph{Subclassification}. The super learner is specified deliberately strongly, following the practical flowchart of \citet{phillips2023}: because the balanced binary outcome puts the effective sample size at $n$, we use $V=\slV$-fold cross-validation stratified on the outcome (and on treatment for the propensity model), a negative-log-likelihood metalearner, and a diverse library spanning the marginal mean, main-effects and penalized (elastic-net) logistic regression, degree-two multivariate adaptive regression splines, random forests, and Bayesian additive regression trees. Because the flexible arms are legitimately strong, any coverage they lose is not an artifact of a weak learner; the cross-fitted diagnostic below then distinguishes a procedure-removable component of the coverage loss from a law-level residual.

\textbf{Diagnostics.} To attribute the breakdown to the law rather than to tuning, three diagnostics accompany the primary arms: a \emph{bridge} TMLE (TMLE-glm) that runs the same targeting step on the main-effects logistic nuisances, with no super learner; a sweep of the propensity truncation bound $g_{\min}\in\{0.01,0.025,0.10\}$ (arms TMLE-gb.01 and TMLE-gb.10 about the default $g_{\min}=0.025$); and a \emph{granularity} coarsening that replaces the covariate by its $K\in\{6,11,23,45\}$ propensity-quantile strata under both mechanisms, holding the super learner fixed. Coarsening changes the covariate resolution together with the confounding information and nuisance difficulty, so an estimator-free \emph{oracle} pass---not coverage recovery alone---isolates the law-level mechanism, computing from the generating construction alone the true overlap floor and single-treatment mass at each resolution (Web Figure~S1). Finally, to separate the induced law from the inferential procedure, we report cross-fitted counterparts of the two super-learner arms---a cross-validated TMLE and a cross-fit augmented IPW \citep{zheng2011,zivich2021,chernozhukov2018}, whose nuisances are fit out-of-fold rather than on the full sample---so that non-cross-fitted (in-sample) nuisance estimation is excluded as an explanation for any coverage loss rather than assumed benign. Software details are in Web Appendix~B.

\textbf{Performance.} Bias, empirical standard deviation, root-mean-square error, and $95\%$ Wald coverage, each with Monte-Carlo error, over $J=\numReps$ replicates; the super-learner/TMLE non-convergence rate is $\approx 0$ throughout, so the coverage results are not a survivorship artifact. Full estimator definitions and the complete cell-by-estimator table (Web Table~S2) are in Web Appendix~B; the software reproduces every number.

\begin{figure}[t]
\centering
\includegraphics[width=\linewidth,alt={Three panels on the right-heart-catheterization
cohort. (a) Interval coverage by estimator under sample- versus generate-treatment: the
super-learner TMLE and augmented IPW fall well below the 0.95 line under sample-treatment
and recover under generate-treatment, while g-computation and IPW stay near 0.95. (b)
RMSE bars, outcome model correct versus wrong, showing the plug-in and doubly-robust
ranking inverting. (c) Absolute bias of the super-learner TMLE rising as the propensity
floor loosens under sample-treatment and flat at zero under generate-treatment.}]{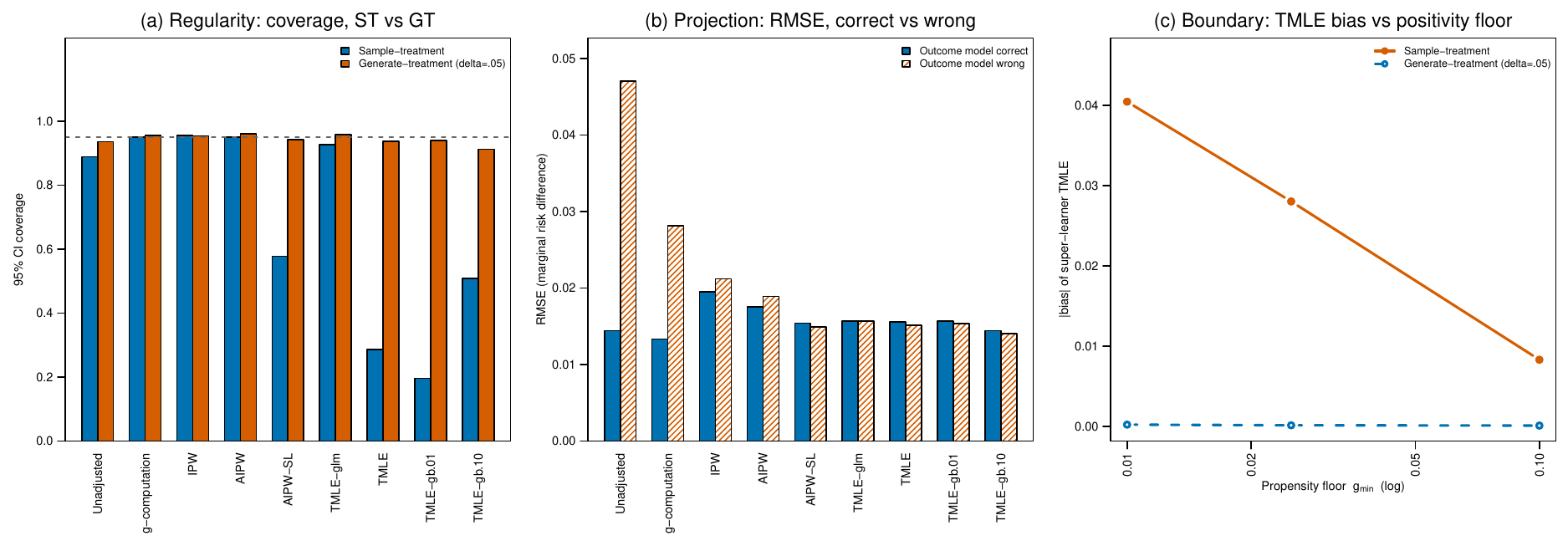}
\caption{Simulation on the right-heart-catheterization cohort ($J=\numReps$).
\textbf{(a)} $95\%$ interval coverage of the marginal risk difference by estimator,
under sample-treatment (left bars) and generate-treatment with a declared floor
(right bars): the super-learner-based efficient estimators (TMLE, AIPW-SL) lose coverage
under sample-treatment and recover under generate-treatment, while the parametric-nuisance
estimators (g-computation, IPW, and the TMLE-glm bridge) stay near-nominal; the dashed
line is the $0.95$ target. \textbf{(b)} RMSE for the marginal risk difference in a regular
generate-treatment world when the analyst's additive outcome model is correct (solid)
versus wrong (hatched): g-computation is efficient when correct and biased when wrong, and
the ranking against AIPW and TMLE inverts even though overlap is intact. \textbf{(c)}
Boundary phenomenon: $|$bias$|$ of the super-learner TMLE against the propensity
truncation floor $g_{\min}$, which grows monotonically under sample-treatment and stays at
zero under generate-treatment. Colour-blind-safe palette; bars differ by fill for
black-and-white reproduction.}
\label{fig:sim}
\end{figure}

\begin{figure}[t]
\centering
\includegraphics[width=\linewidth,alt={Two panels. (a) Law-level: as the covariate is
coarsened into fewer propensity strata, the oracle overlap floor rises off zero and the
single-treatment mass falls. (b) Super-learner TMLE coverage recovers from about 0.29 at
full resolution to about 0.95 as the covariate is coarsened under sample-treatment, while
generate-treatment stays near 0.95 at every resolution.}]{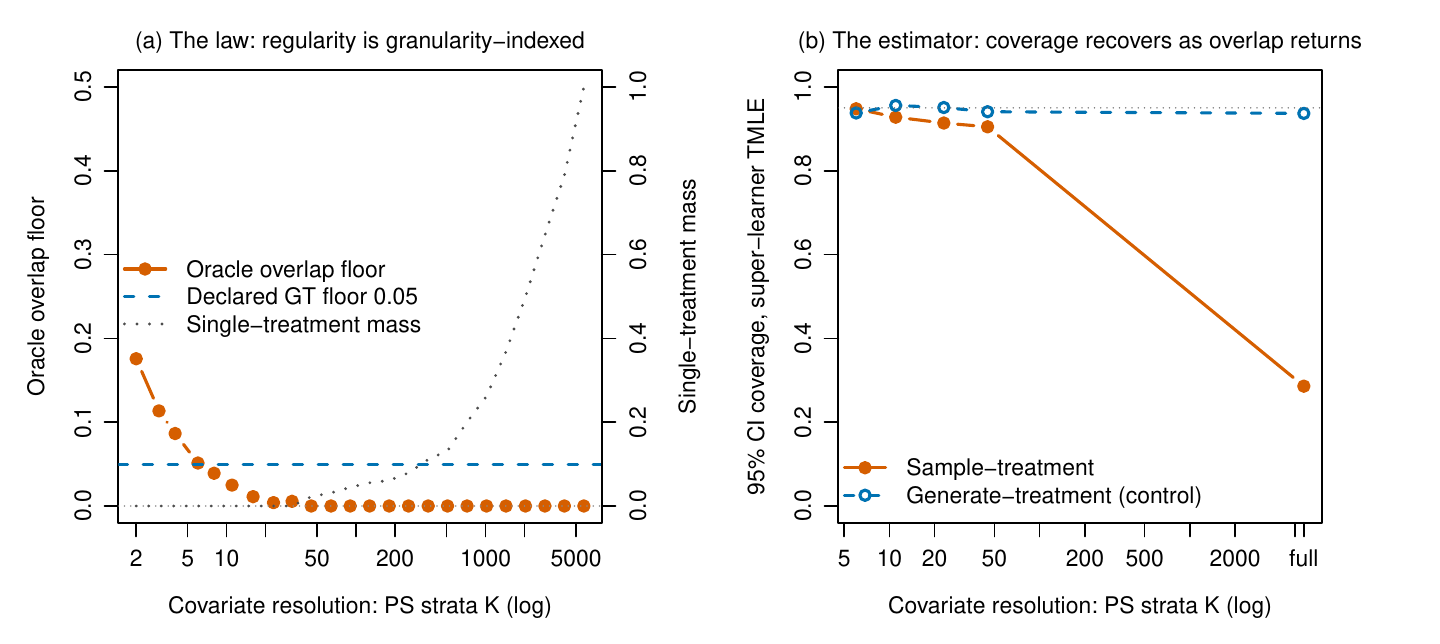}
\caption{The regularity failure is granularity-indexed, shown at the level of the law and
of the estimator. \textbf{(a)} Law only, no estimator: as the covariate is coarsened to
$K$ propensity strata the oracle overlap floor rises off zero and the single-treatment
mass falls---at full resolution every configuration is near-unique and the true propensity
is degenerate by construction. \textbf{(b)} Super-learner TMLE $95\%$ coverage recovers
from the full-resolution value up to nominal as the covariate is coarsened under
sample-treatment, while the generate-treatment control stays regular at every resolution.
The collapse is concentrated at full resolution, tracking the single-treatment mass rather
than the floor alone---the high-dimensional near-determinism of \citet{damour2021}.}
\label{fig:gran}
\end{figure}

\textbf{Results.} Table~\ref{tab:sim} and Figures~\ref{fig:sim}--\ref{fig:gran} come from a
single data-generating engine built on the shared construction of Algorithm~\ref{alg:regular}, with the treatment mechanism as the switched coordinate; the granularity figure extends
the main run to coarser covariate resolutions under both mechanisms.

\textbf{Regularity (Fig.~\ref{fig:sim}a).} Under sample-treatment the coverage
failure falls specifically on the \emph{super-learner-based} efficient estimators: TMLE
and augmented IPW cover the marginal risk difference at \covTmleST{} and \covAipwSlST{}
against the nominal $0.95$, recovering to \covTmleGT{} under a declared-floor
generate-treatment design---all arms scored against the exactly-computable design truth
$\psistar$, which stays well-defined under sample-treatment even where $\psi(\Ptil_n)$
is nonparametrically unidentified (Web Appendix~A). Every \emph{parametric-nuisance} estimator retains
near-nominal coverage under sample-treatment---g-computation \covGcompST{}, IPW
\covIpwST{}, and the correctly-specified TMLE with logistic nuisances \covTmleGlmST{}.
This reproduces \citet{shaw2026}, who report good coverage for their correctly-specified,
parametric estimators under sample-treatment for a common binary outcome; the new
phenomenon is that \emph{data-adaptive} nuisance estimation---standard modern practice,
and exactly what a plasmode is usually built to benchmark---does not share that protection.

\begin{figure}[t]
\centering
\includegraphics[width=0.72\linewidth,alt={Grouped bar chart of 95 percent coverage by
cell for the in-sample and cross-fitted flexible arms. At sample-treatment full
resolution the cross-fitted CV-TMLE and cross-fit AIPW cover higher than their in-sample
counterparts but remain below the 0.95 line; at generate-treatment and at the coarsened
K=6 resolution all arms are near 0.95.}]{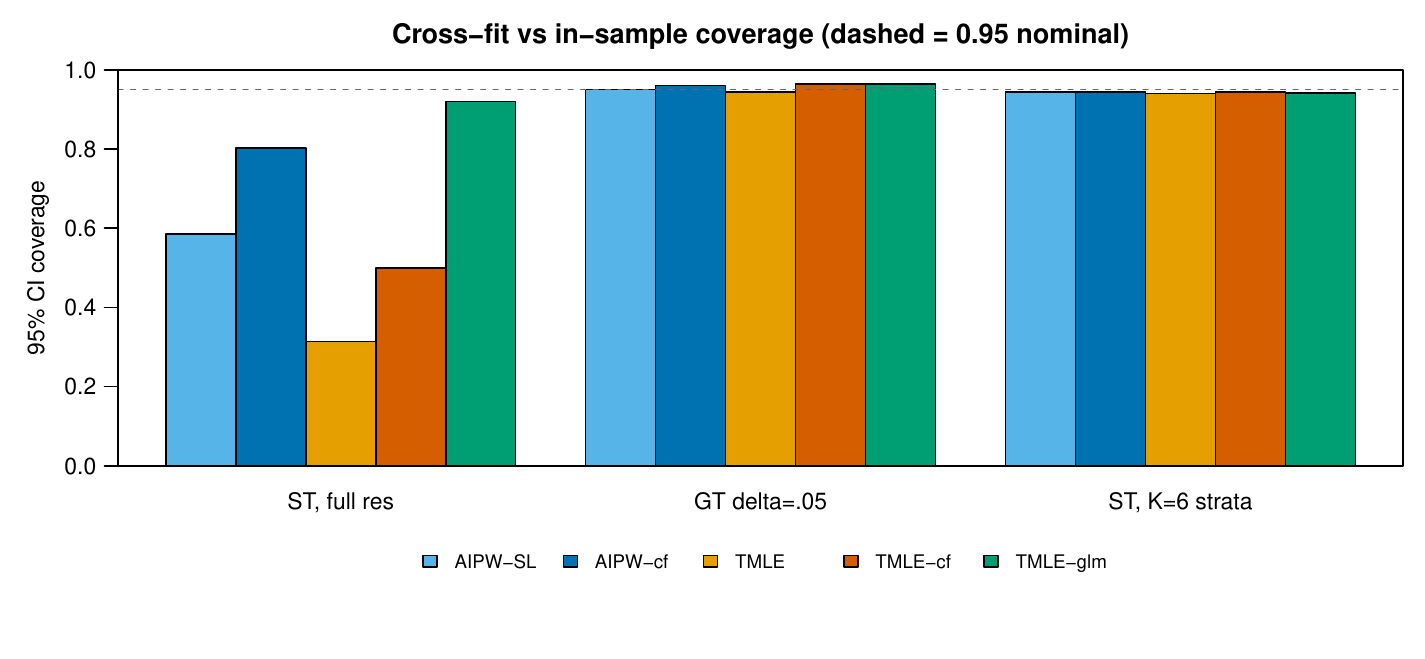}
\caption{Cross-fitting diagnostic ($J=500$): $95\%$ coverage of the flexible arms with
in-sample versus out-of-fold (CV-TMLE, cross-fit AIPW) nuisances. Under sample-treatment
at full resolution, cross-fitting lifts coverage (CV-TMLE
\covTmleCfInsST{}$\to$\covTmleCfST{}, cross-fit AIPW \covAipwCfInsST{}$\to$\covAipwCfST{})
but neither reaches the $0.95$ target (dashed line); under generate-treatment and under
coarsening to $K=6$ strata all arms are near-nominal. The residual sample-treatment
deficit that survives cross-fitting is attributable to the induced law; the part cross-fitting
removes, to the non-cross-fitted inference.}
\label{fig:crossfit}
\end{figure}

\textbf{A law--inference interaction, not a tuning artifact (Fig.~\ref{fig:gran}).} The breakdown is neither a pure property of the law nor a mere tuning artifact, and four results locate it. Three tie it to the induced law's loss of
overlap. First, it is \emph{granularity-indexed}: coarsening the
covariate to $K$ propensity strata lifts the estimator-free oracle overlap floor off zero and drops the single-treatment
mass (Fig.~\ref{fig:gran}a)---a law-level signal, since coarsening also co-varies nuisance
dimension, complexity, and the estimand with overlap, so coverage recovery alone would
not isolate it---and returns super-learner TMLE coverage from \covTmleST{} at full
resolution to \covTmleCoarseST{} at $K=6$, while the generate-treatment control stays
regular at every $K$ (Fig.~\ref{fig:gran}b). Second, the breakdown is monotone in the
positivity bound: as $g_{\min}$ loosens from $0.10$ to $0.025$ to $0.01$ the
sample-treatment bias grows (\biasTmleGbTight{}, \biasTmleGbDef{}, \biasTmleGbLoose{}),
the signature of a boundary phenomenon. Third, an estimator-free law-level diagnostic
confirms the mechanism directly: at full resolution every covariate configuration is
near-unique, so the true propensity is degenerate on $\{0,1\}$ by construction and the
single-treatment mass approaches one (Theorem~\ref{thm:reg}). Fourth, a paired
cross-fitted diagnostic ($J=500$; Fig.~\ref{fig:crossfit}) distinguishes a procedure-removable component of the coverage loss from a law-level residual:
replacing the in-sample super-learner nuisances with out-of-fold fits---CV-TMLE and
cross-fit augmented IPW---lifts sample-treatment coverage from \covTmleCfInsST{} and
\covAipwCfInsST{} to \covTmleCfST{} and \covAipwCfST{}, and cuts the TMLE plug-in bias
from \biasTmleCfInsST{} to \biasTmleCfST{}, yet neither returns to nominal, while both
are near-nominal under generate-treatment (\covTmleCfGT{}) and under coarsening
(\covTmleCfKsix{}). Non-cross-fitted inference is thus a genuine contributor to the
in-sample collapse; the residual sub-nominal coverage that survives cross-fitting under
sample-treatment---and vanishes once overlap is restored---is attributable to the induced
law: because the nonparametric efficiency bound diverges there (Theorem~\ref{thm:reg}), no
efficient-influence-function Wald interval for the super-learner arms can restore nominal
coverage, though the residual magnitude remains estimator-dependent. The parametric-nuisance
estimators are spared not because the law is regular but because a rigid propensity model
cannot resolve that near-determinism, so it does not chase the boundary---their nominal
coverage reflects the rigidity of the working model, not the regularity of the law.

\textbf{Projection (Fig.~\ref{fig:sim}b).} In a regular generate-treatment world the
correctly specified g-computation estimator has the lowest error when its outcome model
matches the generating one (RMSE \rmseGcompOK{} vs AIPW \rmseAipwOK{}, TMLE \rmseTmleOK{}),
and loses once the outcome model is wrong (\rmseGcompWR{} vs \rmseAipwWR{}, \rmseTmleWR{}):
the ranking inverts although overlap is intact. The flexible arms carry the same additive
main-effects submodel in their super-learner library (Table~\ref{tab:design}B), so the
inversion is set by the analyst's injected outcome structure---correct by construction on
the plasmode, generically unavailable at $P_0$---not by estimator capability, and so need
not transfer.

\textbf{Difficulty dial.} Lowering the
declared floor from $\delta=0.10$ to $\delta=0.02$ lowers the effective sample size of the
weights from \essHi{} to \essLo{}---a \emph{declared} law-level difficulty coordinate the
analyst sets and reports (Alg.~\ref{alg:regular}). That estimator performance holds
near-nominal across the dial is itself informative: the declared-floor anchor stays
regular as difficulty tightens.

\begin{table}[t]\centering\small
\caption{Marginal risk-difference performance on the right-heart-catheterization
cohort ($n=\rhcN$, $J=\numReps$ replicates, super learner at $V=\slV$-fold
cross-validation). \emph{Panel A} (regularity): $95\%$ interval coverage under
sample-treatment (ST) versus generate-treatment with a declared floor
$\delta=\floorGT$ (GT). Under ST the super-learner TMLE collapses while the
parametric-nuisance arms---g-computation, IPW, and TMLE with logistic nuisances---stay
near-nominal; all recover under GT. \emph{Panel B} (projection): RMSE in a regular GT
world when the analyst's additive outcome model is correct versus wrong. \emph{Panel C}
(cross-fitting): sample-treatment coverage of the flexible arms with in-sample versus
out-of-fold (CV-TMLE, cross-fit AIPW) nuisances, $J=500$.}
\label{tab:sim}
\begin{tabular}{@{}lrr@{}}
\toprule
\multicolumn{3}{@{}l}{\emph{Panel A. Regularity --- $95\%$ coverage}}\\
Estimator & Sample-treatment & Generate-treatment \\
\midrule
g-computation            & \covGcompST   & \covGcompGT \\
IPW                      & \covIpwST     & \covIpwGT \\
TMLE (super learner)     & \covTmleST    & \covTmleGT \\
AIPW-SL                  & \covAipwSlST  & \covAipwSlGT \\
TMLE-glm (logistic nuisances) & \covTmleGlmST & --- \\
\midrule
\multicolumn{3}{@{}l}{\emph{Panel B. Projection --- RMSE, GT ($\delta=\floorGT$)}}\\
Estimator & Outcome model correct & Outcome model wrong \\
\midrule
g-computation & \rmseGcompOK & \rmseGcompWR \\
AIPW          & \rmseAipwOK  & \rmseAipwWR \\
TMLE          & \rmseTmleOK  & \rmseTmleWR \\
\midrule
\multicolumn{3}{@{}l}{\emph{Panel C. Cross-fitting --- $95\%$ coverage, sample-treatment, full resolution}}\\
Estimator & In-sample & Cross-fit \\
\midrule
TMLE (super learner) & \covTmleCfInsST & \covTmleCfST \\
AIPW-SL              & \covAipwCfInsST & \covAipwCfST \\
\bottomrule
\end{tabular}

\smallskip
\footnotesize\raggedright
Under ST the super-learner/TMLE non-convergence rate is \naTmleST{} versus \naTmleGT{}
under GT. The ST bias is monotone in the propensity truncation bound $g_{\min}$:
\biasTmleGbTight{}, \biasTmleGbDef{}, \biasTmleGbLoose{} at
$g_{\min}=0.10,\,0.025,\,0.01$. Lowering the declared floor from $\delta=0.10$ to
$\delta=0.02$ lowers the effective sample size of the weights from \essHi{} to \essLo{}.
The complete cell-by-estimator table (all cells and arms) is Web Table~S2
(Web Appendix~B); there, propensity-score subclassification is a point estimator without
a Wald interval, so its coverage is not reported (``--''). Panels~A--B use $\numReps$
replicates; the cross-fit diagnostic (Panel~C) uses $J=500$, so its in-sample
sample-treatment baselines (\covTmleCfInsST{}, \covAipwCfInsST{}) differ slightly from
the corresponding $\numReps$-replicate values. Monte-Carlo standard errors are at most
$\approx\!0.016$ for coverage ($\approx\!0.007$ near nominal) and $\approx\!0.0015$ for
bias at $\numReps$ replicates. Cross-fitting (Panel C) lifts ST coverage but not to nominal---both
flexible arms are near-nominal under GT (\covTmleCfGT{}) and coarsening
(\covTmleCfKsix{})---locating the deficit as a law--inference interaction.
\end{table}

\section{What a benchmark can conclude}\label{sec:conclude}

\begin{theorem}[Transfer decomposition: refutation is unconditional, certification requires a transfer premise]\label{thm:falsify}
Write $R_n(k;P)$ for the finite-sample risk of estimator $k$ at law $P$, and
$\sigma(P)$ for the induced ranking. Decompose transfer into (T1) estimand transfer,
(T2) per-estimator difficulty transfer $R_n(k;\Ptil_n)\approx R_n(k;P_0)$, and (T3)
ranking transfer. Then (a) T2 with a gap condition---error below half the smallest
true risk gap---implies T3, but $\sigma$ is discontinuous in the risk vector, so no
``$\Ptil_n$ close to $P_0$'' statement implies T3 near ties; and (b) a single anchor
$\Ptil_n$, when it lies in the class over which a guarantee is quoted, is a genuine
counterexample, so an observed violation there \emph{refutes} that universal claim with
no transfer assumption---so a sample-treatment $\Ptil_n$ with near-unique covariates, lying outside every
strict-overlap class (Theorem~\ref{thm:reg}), refutes only claims asserting validity
without a positivity premise, not the overlap-restricted guarantees augmented weighting
and targeted learning actually carry, which assert nothing at $\Ptil_n$---whereas
\emph{certifying} it requires an external transfer premise linking performance at the two laws---T2 being its canonical risk-closeness instance, though a dominance, uniform-class, or invariance argument would serve equally---which the benchmark data alone cannot establish.
\end{theorem}

This cashes out the title, and it explains why the sharpest uses of empirical
benchmarks are negative: \citet{advani2019} show empirical Monte-Carlo designs can
select an estimator worse than random, and \citet{curth2021} show a benchmark's design
can decide the winner---both are refutations, which need no transfer. Because a
benchmark estimates the failing property by simulation, the refutation holds exactly
for the risk at $\Ptil_n$ and for the $J$-replicate estimate up to a Monte-Carlo error
that is small at $J=\numReps$.

\begin{appliedbox}
Treat a plasmode as a way to \emph{rule estimators out}, never to sign one off. If an
estimator fails on an anchor that genuinely belongs to the class where the estimator claims
to work, that is a real counterexample you can believe, with no assumption that the
benchmark resembles your cohort. Good performance is different: an estimator doing well on a
plasmode does not certify it will do well on your data, because that would need the
benchmark's difficulty to carry over to your population---something the benchmark itself
cannot show. One subtlety when you report a failure: if it comes from a design that has left
the estimator's assumptions behind (for example, a sample-treatment anchor with no overlap),
it does not refute the estimator's guarantee, which never claimed anything there in the
first place. Report which estimators a benchmark breaks and under what conditions, and do
not read a clean pass as a licence to trust.
\end{appliedbox}

\section{Discussion}\label{sec:discussion}

\textbf{What the two axes buy.} Reading plasmode on a projection axis and a regularity
axis separates the two things the recent discussion ran together. The positivity
dispute is the regularity axis, settled by granularity; the older worry that a
parametric outcome model favors parametric estimators is the projection axis, a
statement about ranking transfer, not measurement validity. Together they give the
verdict: a single anchored plasmode refutes universal claims and certifies none.

\textbf{Relation to \citet{shaw2026}.} The two accounts agree on the recommendation
and differ on its reach. \citet{shaw2026} show that sample-treatment biases
propensity-based estimators through a positivity violation and advise
generate-treatment; we reach the same advice and add four things. We \emph{ground}
the mechanism in the high-dimensional overlap geometry of \citet{damour2021}, realized
exactly at finite $n$; we \emph{bound} it, since it is granularity-indexed and absent
under coarse covariates; and---short of coarsening away the real high-dimensional
covariate law, which would restore regularity but forfeit the realism that justifies
plasmode---generate-treatment is the regularity fix we adopt, and it is
\emph{not sufficient}: it repairs positivity but leaves projection untouched
(Figure~\ref{fig:sim}b). Fourth, we \emph{sharpen the estimator scope}. \citet{shaw2026}
establish their result for correctly-specified parametric estimators and report that,
for a common binary outcome, coverage is preserved under sample-treatment---which our
own parametric-nuisance arms (g-computation, IPW, and a correctly-specified TMLE)
reproduce. We show that the failure nonetheless falls on the \emph{data-adaptive}
super-learner estimators that are standard practice for exactly this benchmarking task,
and that cross-fitting the flexible nuisances repairs only part of it---a residual
sample-treatment deficit persists that the parametric-nuisance arms avoid
(Section~\ref{sec:sim})---so a plasmode built to evaluate modern
machine-learning-based estimators is where sample-treatment is most misleading. The
difference is not about their result but about what fixing positivity accomplishes, and
for which estimators the caution bites hardest.

\textbf{Relation to benchmark-design work.} The projection axis formalizes a caution that
recurs across the benchmark-selection literature. \citet{advani2019} show that empirical
Monte-Carlo schemes can select a treatment-effect estimator worse than random;
\citet{curth2021} show that semi-synthetic conditional-effect benchmarks systematically
favour algorithms whose inductive bias matches the generating process; and
\citet{schuler2017} propose selecting a causal method by its accuracy on generators fitted
to the source data---precisely the practice the projection axis diagnoses, since a fitted
generator is a working model that some estimator matches by construction. A parallel
strand raises the \emph{fidelity} of the generator: \citet{athey2024} learn realistic
designs with Wasserstein generative adversarial networks, and \citet{amaranath2026} infer
a data-informed posterior over generator parameters by simulation-based inference. Higher
fidelity is valuable but closes neither axis---a better-fit generator still embeds a
working model (projection), and a more realistic covariate law does not restore the
overlap that a sample-treatment design destroys (regularity). Our contribution is
orthogonal: not a better generator, but two validity axes and a refute/certify principle
that bound what any such benchmark can conclude. In our own concurrent plasmode study of
rare exposures \citep{karim2026rare}, a generate-treatment design with propensity
truncation exhibits exactly the overlap-driven instability the regularity axis predicts.

\textbf{Recommendations.} For an analyst comparing causal estimators with a plasmode,
Table~\ref{tab:practice} draws the rules together. Generate the treatment from a
declared bounded propensity; score every marginal estimator against the standardized
truth recomputed from the process, not the injected coefficient; declare and report
the difficulty coordinates; cross-fit the flexible nuisances (CV-TMLE or cross-fit AIPW),
since in-sample nuisance plugging makes their coverage anti-conservative and confounds the
induced law's contribution; and read the output as a filter, not a warrant---a
plasmode that trips an estimator is informative, but one on which an estimator does
well is not a licence to trust it on the real cohort.

\begin{appliedbox}[How to use this in practice]
To benchmark causal estimators with a plasmode: (1)~\emph{generate} the treatment from a
propensity you bound away from $0$ and $1$ by a floor you choose, rather than carrying the
real treatment through; (2)~compute the benchmark's true effect by standardizing the
outcome model you injected, and score every estimator against \emph{that} number, not
against the coefficient you typed in; (3)~write down and report your difficulty
settings---the floor, the confounding strength, and how complex the outcome surface
is---so a reader knows how hard the comparison was; (4)~for flexible,
machine-learning-based estimators, use cross-fitting (CV-TMLE or cross-fit AIPW), because
fitting the nuisances on the same data makes their intervals too narrow, so their coverage
falls below nominal---cross-fitting lifts it back toward the target (only part of the way
under sample-treatment); and (5)~read the result as a filter---trust the estimators it
\emph{breaks}, and treat a good score as encouragement to look further, not as a guarantee
for your cohort. Table~\ref{tab:practice} ties each step to the result behind it.
\end{appliedbox}

\begin{table}[htbp]\centering\footnotesize
\caption{Recommendations for a plasmode benchmark of causal estimators, each tied to a
result and illustrated on the right-heart-catheterization cohort of
Section~\ref{sec:sim}. A plasmode is a valid filter against estimators, not a warrant for them.}
\label{tab:practice}
\begin{tabular}{@{}>{\raggedright\arraybackslash}p{0.13\textwidth} >{\raggedright\arraybackslash}p{0.25\textwidth} >{\raggedright\arraybackslash}p{0.23\textwidth} >{\raggedright\arraybackslash}p{0.27\textwidth}@{}}
\toprule
Design choice & What goes wrong if it is ignored & Recommendation & Example (RHC cohort) \\
\midrule
Treatment mechanism & Sample-treatment with near-unique covariate patterns drives the propensity to $\{0,1\}$: the law leaves the strict-overlap model, so estimators that target the nonparametric-efficient effect with data-adaptive nuisances (super-learner TMLE/AIPW) lose their coverage and efficiency guarantee, while parametric-nuisance estimators retain coverage only under an unverifiable working-model assumption (Thm~\ref{thm:reg}). & Generate treatment from a declared propensity bounded in $[\delta,1-\delta]$ (Alg.~\ref{alg:regular}). & On RHC the sample-treatment overlap floor collapses to \floorHi{}; declaring a propensity floor $\delta=\floorGT$ bounds the propensity into $[\delta,1-\delta]$ and restores overlap, lifting super-learner TMLE coverage from \covTmleST{} to \covTmleGT{}. \\
\addlinespace
Scoring the truth & The injected coefficient is a conditional odds-ratio; a marginal estimator targets a different number (Prop.~\ref{prop:noncollapse}). & Recompute the marginal truth by standardization from the known process. & The injected conditional log-odds-ratio $\log 2=\numBstar$ becomes a marginal risk difference of $\numMargRD$; scoring against $\numBstar$ would report a large spurious bias. \\
\addlinespace
Difficulty & Overlap, confounding strength, and outcome-model complexity, read off a fit, are neither controlled nor reported. & Impose and report them as declared coordinates (Alg.~\ref{alg:regular}). & The difficulty dial: the weight effective sample size falls from \essHi{} at declared floor $\delta=0.10$ to \essLo{} at $\delta=0.02$. \\
\addlinespace
Flexible-estimator inference & In-sample super-learner nuisances make the efficient estimators' coverage anti-conservative and confound the induced law's contribution with a non-cross-fitted (empirical-process) artifact. & Cross-fit the flexible nuisances (CV-TMLE / cross-fit AIPW; Fig.~\ref{fig:crossfit}). & Cross-fitting lifts sample-treatment coverage (TMLE \covTmleCfInsST{}$\to$\covTmleCfST{}, AIPW-SL \covAipwCfInsST{}$\to$\covAipwCfST{}) but not to nominal---the residual deficit is the induced law's. \\
\addlinespace
Interpretation & A single matched anchor cannot establish class-level reliability (Thm~\ref{thm:falsify}). & Read the benchmark as a falsifier: report the estimators it breaks, not the ones it flatters. & The sample-treatment anchor has left the strict-overlap model $\Mdelta$, so the super-learner TMLE coverage it reports (\covTmleST{}) is an out-of-model measurement, not a refutation of the estimator's in-model guarantee; and its recovery under GT (\covTmleGT{}) licenses no claim about the source cohort. \\
\bottomrule
\end{tabular}
\end{table}

\textbf{What generalizes.} The two axes are not specific to the binary-logistic
plasmode; they are properties of a data-anchored benchmark, and the framework's principles port
beyond the present setting---though the illustration here instantiates them at a single
cohort, estimand, injected effect, and learner library, leaving multi-setting empirical
validation to future work. \emph{Projection} is a property of any benchmark whose truth
is read off a fitted generator: wherever the generating class is a working model some
estimator adopts, that estimator is correct by construction and the measured ranking need
not transfer---the argument uses tangent-space inclusion, not the outcome type.
\emph{Regularity} is a property of any estimand with a support condition: wherever a
guarantee is quoted over a model that requires overlap (more generally, a finite
efficiency bound), a design that drives the induced law to that boundary evaluates
estimators outside the model where their theory holds. The refute/certify asymmetry
(Theorem~\ref{thm:falsify}) is a statement about universal claims and risk vectors, and
applies to any benchmark and any risk criterion. What is specific to the present setting
is the arithmetic, not the principle: the non-collapsibility warning
(Proposition~\ref{prop:noncollapse}) is an odds-ratio phenomenon, and the exact influence
function and $1/\delta$ constants are those of the marginal risk difference. The framework
therefore ports to continuous outcomes, other links, and other estimands; the projection
axis, in particular, extends to any non-causal benchmark whose truth is generator-defined,
while the regularity axis ports wherever a support condition---overlap, or more generally a
finite efficiency bound---is in play. Censoring and time-varying treatment---and with them
the survival case---raise their own positivity questions and are left to future work.

\textbf{Limitations and frontier.} The one open problem is a data-computable diagnostic
for difficulty transfer (T2): the declared-difficulty design controls what is
\emph{imposed} but cannot certify what \emph{transfers} to a named population---an
obstruction analogous to the fundamental limits of structure-agnostic functional
estimation \citep{balakrishnan2023}. A related frontier is the realism--regularity trade-off
curve. Finally, sample-treatment plasmode ships bootstrap copies of real covariate
rows, and the near-unique patterns that drive Theorem~\ref{thm:reg} are exactly the
patterns that make a record identifiable, so the regularity failure and the disclosure
risk are two readings of the same granularity; a formal treatment is left to future
work.

\textbf{Conclusion.} The choice between sample-treatment and generate-treatment is a
choice about regularity; the choice of an outcome-generating model is a choice about
projection; and neither choice buys the one thing a benchmark is often asked for---a
warrant that the estimator ranking it reports is the ranking that would hold on the
source cohort the simulation was built from.

\FloatBarrier
\section*{Supplementary material}
The Web Appendix (a separate document) contains the formal statements and complete
proofs of all results (Web Appendix~A), and the data-generating process for the
illustration together with the complete cell-by-estimator results (Web Appendix~B).

\section*{Acknowledgements}
This research was enabled in part by computational resources provided by Advanced
Research Computing at the University of British Columbia.

\section*{Declarations}
\textbf{Declaration of generative AI use.} During the preparation of this work the
author used AI-based tools to assist with simulation and analysis code, manuscript
editing, and the checking of derivations; the author reviewed and verified all outputs
and takes full responsibility for the content of this article.\\[2pt]
\textbf{Declaration of competing interest.} The author declares no competing interests.\\[2pt]
\textbf{Funding.} This research did not receive any specific grant from funding agencies
in the public, commercial, or not-for-profit sectors.\\[2pt]
\textbf{Data availability.} The illustration uses the publicly available
right-heart-catheterization cohort \citep{connors1996}, distributed for teaching by the
Vanderbilt Department of Biostatistics (\url{https://hbiostat.org/data}). Code and
committed results reproducing every result are at
\url{https://github.com/ehsanx/regmode-code}; the reusable \texttt{regmode} R package is
at \url{https://github.com/ehsanx/regmode}.\\[2pt]
\textbf{CRediT authorship.} M. Ehsan Karim: Conceptualization, Methodology, Formal
analysis, Software, Writing -- original draft, Writing -- review \& editing.

\bibliographystyle{abbrvnat}
\bibliography{ref}

\end{document}